\documentclass[article]{elsarticle}
\makeatletter
\def\ps@pprintTitle{%
	\let\@oddhead\@empty
	\let\@evenhead\@empty
	\def\@oddfoot{\centerline{\thepage}}%
	\let\@evenfoot\@oddfoot}
\makeatother

\usepackage{lineno,hyperref}
\usepackage{threeparttable}
\usepackage{float}
\usepackage{booktabs}
\usepackage{amsmath}
\usepackage{comment}
\newcommand{\comm}[1]{}
\usepackage{color}

\usepackage{amsfonts}
\usepackage{amssymb}
\usepackage{amsthm}

\usepackage{epstopdf}
\usepackage{natbib}

\journal{Computer Methods in Applied Mechanics and Engineering }

\begin{document}

	\begin{frontmatter}
		
		\title{Comment on "Anisotropic peridynamics for homogenized microstructured materials" by Vito Diana, Andrea Bacigalupo, Marco Lepidi and Luigi Gambarotta [Comput. Methods Appl. Mech. Engrg. 392 (2022) 114704]}
		
		\author{Marcin Ma\'zdziarz\corref{mycorrespondingauthor}}
		\ead{mmazdz@ippt.pan.pl}
		
		\address{Institute of Fundamental Technological Research Polish Academy of Sciences, Warsaw, Poland}
		\cortext[mycorrespondingauthor]{Corresponding author}

		
	\end{frontmatter}
	
	%\linenumbers
	
	%\section{The Elsevier article class}
	
Recently, Vito Diana, Andrea Bacigalupo, Marco Lepidi and Luigi Gambarotta \cite{DIANA2022114704} proposed an anisotropic continuum-molecular model based on pair potentials for describing the linear behavior of 
periodic heterogeneous {Cauchy} materials in the framework of peridynamic theory. 

Unfortunately, \comm{the consequences resulting from the properties of the {Cauchy} material were ignored,
	and} the proposed model for an isotropic material may have zero or even negative shear stiffness and this will be the case for \emph{Silicene} or \emph{Germanene}, for example. In addition, 2D symmetries were confused with 3D symmetries, and hence the limits for the Poisson's ratio. 

In Ref.\cite[Eq.38]{DIANA2022114704} two elastic quadratic potentials of pairwise axial (${w}_{s}$) and shearing deformation measures (${w}_{\gamma}$) in the special case of isotropic materials reduce to: 

\comm{\begin{align}
		\begin{split}
			{w}_{s}=\frac{1}{2}s^2\left|\mathbf{\xi}\right|k_n=6\pi{h}{\delta}^3 s^2 \left|\mathbf{\xi}\right|\left(C_{1111}-C_{1212}  \right),\\
			{w}_{\gamma}=\frac{1}{2}{\gamma}^2\left|\mathbf{\xi}\right|k_t=6\pi{h}{\delta}^3{\gamma}^2 \left|\mathbf{\xi}\right|\left(3C_{1212}-C_{1111}\right).
		\end{split}
		\label{eqn:w}
\end{align}}

\begin{align}
	%	\begin{split}
		{w}_{s}=\frac{1}{2}s^2\left|\mathbf{\xi}\right|k_n=6\pi{h}{\delta}^3 s^2 \left|\mathbf{\xi}\right|\left(C_{1111}-C_{1212}  \right).
		%	\end{split}
	\label{eqn:w1}
\end{align}

\begin{align}
	%	\begin{split}
		{w}_{\gamma}=\frac{1}{2}{\gamma}^2\left|\mathbf{\xi}\right|k_t=6\pi{h}{\delta}^3{\gamma}^2 \left|\mathbf{\xi}\right|\left(3C_{1212}-C_{1111}\right).
		%	\end{split}
	\label{eqn:w2}
\end{align}

\comm{Let us analyze the behavior of these potentials for the {Cauchy} material. For fully anisotropic ordinary material the number of independent components of four-rank elasticity tensor $C_{ijkl}$ reduces to 21 in 3D and to 6 in 2D, see \cite{Mazdziarz2019}. This reduction is due to the so-called minor ($C_{ijkl} = C_{jikl} = C_{ijlk}$) and major ($C_{ijkl} = C_{klij}$) symmetry of this tensor. But for a {Cauchy} material, i.e. based upon the hypothesis of central forces acting between any couples of molecules/atoms/particles, we have additional symmetry: ($C_{ijkl} = C_{ikjl}$), see \cite{Campanella1994}. The elasticity tensor $C_{ijkl}$ becomes hence a fully symmetric tensor and this additional symmetry reduces the number of independent components to 15 in 3D and to 5 in 2D (probably an omission of this fact by the Authors resulted in faults in the shearing potential ${w}_{\gamma}$ and in benchmarks).  Let us now consider the case of isotropy.} The elasticity tensor for an ordinary 2D material with isotropic symmetry has a representation in the Voigt notation:

\begin{align}
	\centering
	%\left[
	%\begin{array}{c}
	{C}_{ijkl}
	%\end{array}
	%\right]
	\rightarrow\left[
	\begin{array}{ccc}
		{{C}_{1111}} & {{C}_{1122}} & 0 \\
		{{C}_{1122}} & {{C}_{1111}} & 0 \\
		0 & 0 & C_{1212}  \\
	\end{array}
	\right], 
	\label{eqn:Hisotropy}
\end{align}
where $C_{1212}=\dfrac{C_{1111}-C_{1122}}{2}$, hence the number of independent components reduces here to 2. \comm{For a {Cauchy} material, an additional condition ($C_{1122} = C_{1212}$) implies that the elasticity tensor for a {Cauchy} 2D material reduces to:} We can therefore write this as:
\begin{align}
	\centering
	%\left[
	%\begin{array}{c}
	{C}_{ijkl}
	%\end{array}
	%\right]
	\rightarrow\left[
	\begin{array}{ccc}
		{{C}_{1111}} & {{C}_{1122}} & 0 \\
		{{C}_{1122}} & {{C}_{1111}} & 0 \\
		0 & 0 & \dfrac{C_{1111}-C_{1122}}{2}  \\
	\end{array}
	\right]=
	\left[
	\begin{array}{ccc}
		{{C}_{1111}} & {{C}_{1111}-2{C}_{1212}} & 0 \\
		{{C}_{1111}-2{C}_{1212}} & {{C}_{1111}} & 0 \\
		0 & 0 & C_{1212}  \\
	\end{array}
	\right]. 
	\label{eqn:Cisotropy}
\end{align}
\comm{Taking into account the further condition of isotropy, i.e.  $C_{1212}=\dfrac{C_{1111}-C_{1122}}{2}$, we obtain:
	\begin{align}
		\centering
		%\left[
		%\begin{array}{c}
		{C}_{ijkl}
		%\end{array}
		%\right]
		\rightarrow\left[
		\begin{array}{ccc}
			{{C}_{1111}} & {\dfrac{C_{1111}}{3}} & 0 \\
			{\dfrac{C_{1111}}{3}} & {{C}_{1111}} & 0 \\
			0 & 0 & {\dfrac{C_{1111}}{3}}  \\
		\end{array}
		\right], 
		\label{eqn:CHisotropy}
	\end{align}
	and hence the number of independent components reduces here only to 1.}
The mechanical stability condition requires the elasticity tensor $C_{ijkl}$ to be positive definite, see \cite{Mazdziarz2019}. This imposes conditions on the elastic constants and for any isotropic two-dimensional \comm{ordinary} material (2D) means:

\begin{align}
	%	\begin{split}
		C_{1212}>0,  
		%		C_{1111}>C_{1212}.
		%	\end{split}
	\label{eqn:cond1}
\end{align}
\begin{align}
	%	\begin{split}
		%		C_{1212}>0, \\ 
		C_{1111}>C_{1212}.
		%	\end{split}
	\label{eqn:cond2}
\end{align}

We see that the condition \ref{eqn:cond2} implies positivity of	\emph{{w$_s$}} in Eq.\ref{eqn:w1} but for \\ {${C_{1111}}>3C_{1212}$} \emph{{w$_{\gamma}$}} in Eq.\ref{eqn:w2} is negative, which is not acceptable as the stiffness cannot be negative. 

This is not some contrived, theoretical case, e.g. for $Silicene$  $C_{1111}$=68.3\,N/m, $C_{1212}$=22.5\,N/m, what means that $\left(3C_{1212}-C_{1111}\right)$\,=\,-0.8\,N/m, see \cite{Wang2019}.

\comm{It is known that for any isotropic {Cauchy} material, the Lam\'{e} coefficients of elasticity satisfy the condition: $\lambda=\mu$, see \cite{Campanella1994} and therefore the Poisson's ratio $\nu=\frac{1}{4}$.
	
	Rewriting the equation \ref{eqn:w} for the isotropic {Cauchy} material we obtain:
	\begin{align}
		{w}_{s}=6\pi{h}{\delta}^3 s^2 \left|\mathbf{\xi}\right|\left(\frac{2}{3}C_{1111}\right), {w}_{\gamma}=6\pi{h}{\delta}^3{\gamma}^2 \left|\mathbf{\xi}\right|\left(0\right),
		\label{eqn:w3}
	\end{align}
	hence $k_n=\frac{2}{3}C_{1111}$ and $k_t=0$. A body subjected to simple shear, therefore, has no stiffness. }

In Ref.\cite[Section 5.1, P.16]{DIANA2022114704} is written: "\emph{... leads to the effective fourth order elasticity tensor $\mathbb{C}$ having the elasticities of the two-dimensional cubic symmetry ...}". No such symmetry exists for 2D materials, there is isotropy, tetragonal, orthotropy and anisotropy, see \cite{Mazdziarz2019}.

\comm{In Ref.\cite[Section 5.2, Tab.1, P.24]{DIANA2022114704} "\emph{Elastic constants of the anisotropic material considered}" are given. Since 2D {Cauchy} material is considered in the paper, $C_{1122}$ must equal $C_{1212}$, and they are not.}

In Ref.\cite[Section 5.1, P.17]{DIANA2022114704} is written: "\emph{... representative extreme values $\nu_a$ $\approx$ 1/2 (e.g. 0.49), ...}". For any isotropic two-dimensional \comm{ordinary} material (2D) the correct limits for the Poisson's ratio, resulting from the positive definiteness of the stiffness tensor, are $-1{\leq}{\nu}{\leq}1$, see \cite{Wojciechowski2003}.

	%{\cite{Hehl2002} "{The Cauchy Relations in Linear Elasticity Theory}"
		%\cite{Ostoja2002} "{Lattice models in micromechanics }"}

	%\section*{References}
	
	\bibliography{CMAME_2022bib}

\begin{thebibliography}{1}
\expandafter\ifx\csname url\endcsname\relax
  \def\url#1{\texttt{#1}}\fi
\expandafter\ifx\csname urlprefix\endcsname\relax\def\urlprefix{URL }\fi
\expandafter\ifx\csname href\endcsname\relax
  \def\href#1#2{#2} \def\path#1{#1}\fi

\bibitem{DIANA2022114704}
V.~Diana, A.~Bacigalupo, M.~Lepidi, L.~Gambarotta, {Anisotropic peridynamics
  for homogenized microstructured materials}, Computer Methods in Applied
  Mechanics and Engineering 392 (2022) 114704.
\newblock \href {https://doi.org/10.1016/j.cma.2022.114704}
  {\path{doi:10.1016/j.cma.2022.114704}}.

\bibitem{Mazdziarz2019}
M.~Ma{\'{z}}dziarz, {Comment on `The Computational 2D Materials Database:
  high-throughput modeling and discovery of atomically thin crystals'}, 2D
  Materials 6~(4) (2019) 048001.
\newblock \href {https://doi.org/10.1088/2053-1583/ab2ef3}
  {\path{doi:10.1088/2053-1583/ab2ef3}}.

\bibitem{Wang2019}
Z.-Q. Wang, T.-Y. L{\"u}, H.-Q. Wang, Y.~P. Feng, J.-C. Zheng, Review of
  borophene and its potential applications, Frontiers of Physics 14~(3) (2019)
  33403.
\newblock \href {https://doi.org/10.1007/s11467-019-0884-5}
  {\path{doi:10.1007/s11467-019-0884-5}}.

\bibitem{Wojciechowski2003}
K.~W.~Wojciechowski, {Remarks on “Poisson Ratio beyond the Limits of the
  Elasticity Theory”}, Journal of the Physical Society of Japan 72~(7) (2003)
  1819--1820.
\newblock \href {https://doi.org/10.1143/JPSJ.72.1819}
  {\path{doi:10.1143/JPSJ.72.1819}}.

\end{thebibliography}
	
\end{document}